\documentclass[prd,aps,amsfonts,superscriptaddress,nofootinbib,longbibliography,notitlepage,twocolumn]{revtex4-1}

\usepackage{graphicx}
\usepackage{xcolor}
\usepackage{rotating}
\usepackage{amsmath,amssymb,graphics,amsthm,isomath}

\usepackage[colorlinks=true, urlcolor=violet, linkcolor=blue, citecolor=red, hyperindex=true, linktocpage=true]{hyperref}
\usepackage[capitalise,compress]{cleveref}

\newcommand{\revision}[1]{{\color{black} #1}}

\makeatletter
\renewcommand{\p@subsection}{}
\renewcommand{\p@subsubsection}{}
\makeatother

\usepackage{xcolor}
\usepackage{mathtools}

\usepackage{dsfont}

\begin{document}
\title{Fracton hydrodynamics}

\author{Andrey Gromov}
\affiliation{Brown Theoretical Physics Center \& Department of Physics,
Brown University, 182 Hope Street, Providence, RI 02912, USA}

\author{Andrew Lucas}
\email{andrew.j.lucas@colorado.edu}
\affiliation{Department of Physics and Center for Theory of Quantum Matter, University of Colorado, Boulder, CO 80309, USA}

\author{Rahul M. Nandkishore}
\affiliation{Department of Physics and Center for Theory of Quantum Matter, University of Colorado, Boulder, CO 80309, USA}

\begin{abstract}
We introduce new classes of hydrodynamic theories inspired by the recently discovered fracton phases  of  quantum  matter. Fracton phases are characterized by elementary excitations (fractons) with restricted mobility. The hydrodynamic theories we introduce describe thermalization in systems with fracton-like mobility constraints,  including fluids where charge and dipole moment are both locally conserved, and fluids where charge is conserved along every line or plane of a lattice.  Each of these fluids is  subdiffusive, and constitutes a new universality class of hydrodynamic behavior.  There are infinitely many such classes, each with distinct subdiffusive exponents, all of which are captured by our formalism. Our framework naturally explains recent results on dynamics with constrained quantum circuits, as well as recent experiments with ultracold atoms in tilted optical lattices.  We identify crisp experimental signatures of these novel hydrodynamics, and explain how they may be realized in near term ultracold atom experiments. 

\end{abstract}

\date{July 22, 2020}

\maketitle

\section{Introduction}
Hydrodynamics describes a universal effective theory for many-body dynamics and thermalization, whether or not the microscopic dynamics is classical or quantum.  Indeed, hydrodynamic behavior for fluids whose microscopic character is intrinsically quantum mechanical has experimentally been observed in superfluid liquid helium \cite{PhysRevLett.17.74}, quark-gluon plasma \cite{Shuryak_2009}, cold atomic gases \cite{Cao_2010}, and electron \cite{de_Jong_1995,Crossno_2016,Bandurin_2016,Moll_2016,Krishna_Kumar_2017,Gooth_2018,sulpizio,jenkins2020imaging} and phonon \cite{Martelli_2018} liquids in solid-state devices. 

This paper develops the novel hydrodynamics of an entirely new kind of quantum matter, in which the elementary excitations are \emph{fractons} -- particles which exhibit constrained dynamics, being either unable to move in isolation, or able to move only in certain directions.  First discovered in exactly solvable lattice models \cite{chamon2005quantum, haah2011local, vijay2015new}, fracton phases are now at the frontier of multiple areas of theoretical physics.  In quantum field theory, the existence of these phases challenges the canonical paradigm that low energy effective theories describe phases of matter. In quantum information, the immobility of fractons may lead to robust quantum memory. And in condensed matter physics, fracton phases are changing our understanding of what properties {\it can} be exhibited by a phase of matter.  Inspired by these surprising and challenging questions, an enormous effort has been made in recent years to study and classify the novel quantum phases of matter containing fractons \cite{prem2017emergent, prem2018cage, prem2018pinch, slagle2018symmetric, 2019Song, slagle2017fracton, slagle2017quantum, slagleXcube2017, shirley2018foliated, slagle2018foliated, pretko2017emergent, pretko2017higher, devakul2018correlation, devakul2018fractal, you2018subsystem, you2018symmetric, you2019fractonic, weinstein2018absence, wang2019higher, seiberg2019field, aasen2020topological, ma2017fracton, yuan2019fractonic, ma2018topological, schmitz2018recoverable, ma2018higher, moudgalya, sous2019fractons}; see also the review articles \cite{nandkishore2018fractons,pretko2020fracton}.
 Nevertheless, because these phases largely lie outside conventional frameworks, many basic questions remain open. For example, what are the transport coefficients of finite temperature fracton matter? How do these models relax to thermal equilibrium, if at all?

While local mobility constraints can lead to very long relaxation times for equilibration \cite{PremHaahNandkishore} and can even produce localization in certain subspaces \cite{pai2019localization, khemani2019local, 2020PollmannFragmentation}, typical initial states in most fracton systems can reach local equilibrium. The late time relaxation to equilibrium should admit a hydrodynamic description.  The qualitative nature of the hydrodynamics will depend on the type of local conservation laws present in the system, and will generally look completely distinct from usual hydrodynamics, such as the Navier-Stokes equations, or Fick's diffusion law.  
%This paper describes the hydrodynamics that emerges in systems with the symmetries and conservation laws of fracton phases. 
Importantly, while the behavior of fractonic phases is highly sensitive to details of the regularization, their hydrodynamic description is sensitive only to symmetries.  As such, multiple fracton phases fall into the same hydrodynamic universality class. We also provide examples where microscopic models that do not have fracton excitations nevertheless fall into the same hydrodynamic universality classes as fluids of fractons.  
 
For example, in certain fracton models, the restricted mobility of excitations can formally be understood as a consequence of the fact that the many-body dynamics conserves not only the total charge density of fractons, but also the total dipole moment (or higher multipole moment) associated with this charge \cite{pretko2017generalized, pretko2017subdimensional, gromov2018towards}. Systems with such conservation laws naturally couple to symmetric tensor gauge theories \cite{xu2006novel, xu2010emergent, rasmussen2016stable}, just like theories of classical or quantum elasticity \cite{kleinert1982duality, kleinert1983double, kleinert1983dual,bijlsma1997collective, Beekman2017rev, Beekman2017,pretko2018fracton, gromov2019chiral, pai2018fractonic, radzihovsky2020fractons, gromov2019duality, gromov2020duality}.  Other fracton models are understood by duality to systems with subsystem symmetries \cite{vijay2016fracton}.  Whether invoking multipole conservation laws or subsystem symmetry, fracton hydrodynamics is properly understood as a set of unusual conservation laws involving the higher rank current operators which are sourced by the higher rank gauge fields.  The consistency of hydrodynamics in the presence of these background gauge fields imposes non-trivial constraints and can lead to slow, subdiffusive thermalization. 

\section{Dipole conservation} 
We begin with the simplest non-trivial example of a fracton fluid: a chaotic local many-body system in $d$ spatial dimensions where each individual term in the Hamiltonian conserves every component of dipole moment:  for any constants $a$ and $b_i$, if $\rho(\mathbf{x},t)$ denotes the density of a locally conserved U(1) charge: \begin{equation}
    \frac{\mathrm{d}}{\mathrm{d}t} \int \mathrm{d}^d\mathbf{x} \; (a+b_ix^i) \rho(\mathbf{x},t) = 0. \label{eq:dipole}
\end{equation}
For simplicity, we assume this is the only conservation law in the system;  this minimal example is sufficient to capture our key results.   Such systems can be realized in chaotic Floquet circuits on a lattice \cite{khemani2019local, khemanihermelenandkishore}, with gate range sufficiently large to ensure thermalization of typical states.   Alternatively, we may consider energy-conserving dynamics in a theory with charge conjugation symmetry at zero density, where $\rho$ decouples from the energy fluctuations within linear response. We focus on the long wavelength, long time limit of such systems, which is described by a continuum effective theory: hydrodynamics. For simplicity we also assume \revision{microscopic} time reversal symmetry and rotational invariance, and emergent homogeneity in space and time.  Since $\rho(\mathbf{x},t)$ is the only locally conserved quantity, our hydrodynamic theory will be a single equation of motion for $\rho$.  The hydrodynamic modes are long wavelength fluctuations in $\rho$, on length scales much larger than the microscopic lattice scale, and as a consequence \revision{the local dipole density is not a hydrodynamic degree of freedom.  After all, a local ``bound" dipole pair can be created or destroyed by the motion in space of a single charge.  The absence of a dipole density as an emergent hydrodynamic degree of freedom is analogous to the absence of angular momentum density as a hydrodynamic mode in a conventional rotationally invariant fluid \cite{usnonabelian}.}

In general, $\rho$ will \emph{not} represent an electrical charge; it could correspond to the number of atoms, or the number of spins pointing up, or could be some emergent quantity in a strongly correlated system.  Nevertheless, analogies with electromagnetism are useful \cite{pretko2017generalized, pretko2017higher} and will lead to the correct analytical framework for hydrodynamics.

 Following Landau's canonical framework, it is tempting to write down $\partial_t \rho = -\partial_i J_i$, and look for functions $J_i(\rho, \partial_j\rho, \partial_j\partial_k\rho,\ldots)$ for which (\ref{eq:dipole}) is obeyed, and which contain the fewest spatial derivatives.   This method leads to the ordinary Fick's Law: $J_i \approx -D\partial_i \rho$, since $\partial_t\rho = D \nabla^2 \rho$, and upon integrating by parts two times in (\ref{eq:dipole}) we indeed obtain zero.  In fact, the canonical diffusion equation in the infinite plane conserves both net charge and net dipole moment: this can be seen by noting that the diffusion equation is linear, and its Gaussian kernel conserves total charge and dipole.

However, this argument is \emph{wrong}.  There are two hints why.  Firstly, for generic boundary conditions, the ordinary diffusion equation does not conserve dipole moment in a closed but finite box.  Secondly, the Einstein relation suggests that a finite diffusion constant $D$ implies a finite conductivity:  a uniform electric field would excite a charge current.  Yet microscopically, the force on a dipole $\mathbf{p}$ is $\mathbf{F} = (\mathbf{p}\cdot \nabla)\mathbf{E}$, and this vanishes in a uniform field. \revision{Since a charge is not by itself mobile, and the mobile excitation (the dipole) feels no force from a constant electric field, we conclude that the conductivity -- as well as the diffusion constant -- must vanish.}

A correct derivation of hydrodynamics, which will resolve these two puzzles, requires a more careful approach.  Charge conservation relies on a global U(1) symmetry.  The low energy degree of freedom is the phase $\phi$ associated with global U(1) transformations. Local changes in U(1) phase $\phi$ source the charge density $\rho$, along with the associated current.  In a model with this symmetry, the effective Lagrangian reads \begin{equation}
    \mathcal{L} = c_1 (\partial_t \phi)^2 - c_2 (\partial_i \partial_j \phi)^2 - c_3 (\nabla^2 \phi)^2 + \cdots.  \label{eq:dipoleL}
\end{equation}
(\ref{eq:dipole}) implies that $\mathcal{L}$ must be invariant under $\phi \rightarrow \phi + a + b_ix^i$, which is why (\ref{eq:dipoleL}) only contains higher derivative terms in space. Similar actions arose in the study of plaquette models \cite{Paramekanti_2002,PhysRevB.77.134449}.   We now couple this theory to external sources, i.e. background gauge fields, following \cite{gromov2018towards}:  writing $\partial_t\phi  \rightarrow \partial_t\phi - A_t $, $\partial_i\partial_j \phi \rightarrow \partial_i\partial_j\phi - A_{ij} $.  The gauge field is not a 1-form, but a mixed rank object $(A_t, A_{ij})$.  

It is helpful to flip the picture around.  $(A_t, A_{ij})$ are background gauge fields which couple to $(\rho, J_{ij})$, where $\rho$ is the conserved charge density and $J_{ij}=J_{ji}$ is a symmetric rank-2 tensor.  As usual, correlation functions of $\rho$ and $J_{ij}$ are generated by \begin{equation}
    Z[A]= \left\langle \exp\left[\mathrm{i}\int \mathrm{d}^{d+1}x ( A_t \rho + A_{ij}J^{ij})\right]\right\rangle.
\end{equation}
The local U(1) conservation law implies \begin{equation}
    Z[A_t, A_{ij}] = Z[A_t + \partial_t \Phi , A_{ij} - \partial_i\partial_j \Phi] 
\end{equation}
where $\Phi$ is a classical background gauge transformation. Taking a functional derivative $\delta Z/\delta \Phi = 0$, we obtain the Ward identity for charge conservation: \begin{equation}
    \partial_t \rho + \partial_i \partial_j J_{ij} = 0. \label{eq:dipolecons}
\end{equation}
The conserved current for this theory is not a vector, but the tensor $J_{ij}$ which counts the flux of dipoles in direction $x_i$ through the $x_j$-plane.  The ``conventional charge current" that counts the flux of charged objects through a surface is given by \begin{equation}
    J_i = \partial_j J_{ij}. \label{eq:JiJij}
\end{equation}
But $J_{ij}$, not $J_i$, is fundamental.

It remains to relate $J_{ij}$ to $\rho$.  
The hydrodynamic paradigm states that one should write $J_{ij}$ as a Taylor expansion in spatial derivatives of $\rho$ ($\rho$, $\partial_i\rho$, $\partial_i\partial_j\rho$, etc.) and that the dominant terms have the fewest derivatives.  The simplest possibility appears to be $J_{ij} = f(\rho)\delta_{ij}$, which leads us right back to Fick's law of diffusion.  At this point, our formal detour immediately pays off: since the current operators are $J_{ij}$ and not $J_i$, $f(\rho)$ is the expectation value of an operator in thermal equilibrium; therefore, under time reversal $f(\rho) \rightarrow f(\rho)$.  However, according to (\ref{eq:dipolecons}), $f(\rho) \rightarrow -f(\rho)$ under time reversal.  These two conditions enforce $f(\rho)=0$, and therefore we must include spatial derivatives in $J_{ij}$: at leading order, we find \begin{equation}
    J_{ij} = -B_1 \left(E_{ij} - \partial_i\partial_j\mu\right) - B_2 \delta_{ij}\left(E_{kk} - \partial_k\partial_k\mu\right) \label{eq:Jijfinal}
\end{equation}
where $\mu \approx \rho/\chi$ for small fluctuations (here $\chi$ is a thermodynamic coefficient), and $E_{ij} = -\partial_t A_{ij} +\partial_i\partial_j A_t$ is the higher-rank electric field which couples to the fluid.  Combining (\ref{eq:dipolecons}) and (\ref{eq:Jijfinal}) we see that the decay of the local density is clearly subdiffusive: modes at wave number $k$ decay at rate $Bk^4/\chi$, where $B=B_1+B_2$.  

We may carefully justify (\ref{eq:Jijfinal}) using the abstract formalism of \cite{Crossley:2015evo}; see also \cite{Haehl:2015foa,Jensen:2017kzi}.  Yet we also know (\ref{eq:Jijfinal}) must be correct because the current $J_{ij}$ must not locally distinguish ``electric sources" $E_{ij}$ from electrochemical potential gradients $\partial_i\partial_j\mu$.  Moreover, if we include \emph{local} stochastic Gaussian noise in (\ref{eq:Jijfinal}), the fluctuation-dissipation theorem is only obeyed if $J_{ij}$ is given by (\ref{eq:Jijfinal}) at leading order in derivatives: see Appendix \ref{app:multipole} for details.  Lastly, we may relate $E_{ij}$ to the ``physical" electric field:  \cite{pretko2017emergent}  \begin{equation}
    2E_{ij} = \partial_i E_j + \partial_j E_i,  \label{eq:Eij}
\end{equation}
and so as expected, a dipole flux arises only in electric field gradients.

\section{Higher multipole conservation}
It is straightforward to generalize the above discussion to higher multipole moments.  For simplicity, let us consider a theory where \begin{equation}
    \frac{\mathrm{d}}{\mathrm{d}t} \int \mathrm{d}^d\mathbf{x} \; (a+a_ix^i+\cdots + a_{i_1\cdots i_n}x^{i_1}\cdots x^{i_n}) \rho = 0, \label{eq:multipole}
\end{equation}
namely all multipoles up to order $n$ are conserved.  The background gauge field becomes $(A_t, A_{i_1\cdots i_{n+1}})$, the charge conservation equation reads $\partial_t\rho + \partial_{i_1}\cdots \partial_{i_{n+1}}J_{i_1\cdots i_{n+1}} = 0$, and (\ref{eq:Jijfinal}) generalizes to $J_{i_1\cdots i_{n+1}} = (-1)^n B^\prime (E_{i_1\cdots i_{n+1}} - \partial_{i_1}\cdots \partial_{i_{n+1}}\mu) + \cdots$ up to other tensor structures.   In the absence of source fields and noise, \begin{equation}
\partial_t\rho + B^\prime (-\nabla^2)^{n+1} \rho = 0;   \label{eq:multipolediffusion}
\end{equation}
density modulations at wavelength $\lambda$ relax in time $\tau \sim \lambda^{2+2n}/B^\prime$.

\revision{
We show in Appendix \ref{app:multipole} that these effective theories of multipole-conserving hydrodynamics are universal fixed points under renormalization group flow.  All nonlinear corrections and nonlinearities in noise beyond the linear response theory we have described are irrelevant.
}

\section{Transport}
Generalizing (\ref{eq:JiJij}) and (\ref{eq:Eij}), a theory with the first $n$ multipoles conserved naturally couples to $n$ derivatives of the electric field:  the $n$-pole flux $J_{i_1\cdots i_{n+1}} \sim \partial_{(i_1}\cdots \partial_{i_n}E_{i_{n+1})}$.  The analogue of the Ohmic resistivity is the prefactor $B^\prime$ of this proportionality.  Suppose, however, that we wish to measure the flux of charge (namely the ``conventional" current) that flows in response to an \emph{electric potential difference}.  For simplicity, consider an (effectively) one dimensional system.   Since the current $J =  B^\prime (-\partial_x^{2})^nE \propto \partial_x^{2n+1}V$ where $V$ is the electric potential, a constant current flows when $V \propto x^{2n+1}$.  A simple calculation then implies that Ohm's Law holds, but the electrical resistance $R$ of a long wire of length $L$ and cross-sectional area $A$ becomes \begin{equation}
    R = \frac{L^{2n+1}}{(2n+1)!B^\prime A}.
\end{equation}The unusual length dependence of $R$ is a striking prediction of subdiffusion.

\section{Subsystem symmetries}
We now turn to a different example.  Consider a theory where the charge density is conserved on every row and column of a two-dimensional square lattice.  In the continuum limit, \begin{equation}
    \frac{\mathrm{d}}{\mathrm{d}t} \int\limits_{y=a} \mathrm{d}x \rho = \frac{\mathrm{d}}{\mathrm{d}t} \int\limits_{x=b} \mathrm{d}y \rho  = 0. \label{eq:squareconservation}
\end{equation}
Here, there is a \emph{single} current operator $J_{xy}$; the background gauge field is $(A_t, A_{xy})$; gauge invariance demands that $Z[A_t,A_{xy}]=Z[A_t+\partial_t \Phi,A_{xy} + \partial_x \partial_y\Phi]$, which leads to the Ward identity and continuity equation \begin{equation}
\partial_t \rho + \partial_x\partial_y J_{xy}=0.    
\end{equation} 
The hydrodynamic theory compatible with gauge invariance and the fluctuation-dissipation theorem corresponds to the choice $J_{xy} = -B(E_{xy}-\partial_x\partial_y\mu)$, which leads to the subdiffusive equation \begin{equation}
    \partial_t \rho = -C \partial_x^2\partial_y^2 \rho. \label{squaresubsystem}
\end{equation}

In an ordinary fluid, reducing the lattice point group symmetry simply includes more complicated tensor structures in the hydrodynamic equations, while leaving their general form unchanged \cite{Cook_2019,varnavides2020generalized}.  Yet in models with subsystem symmetry, the microscopic lattice plays a critical role in how subdiffusive the dynamics can be.    For example, consider a triangular lattice: in the continuum charge must be conserved along any line of the form $x=a$, $\sqrt{3}y\pm x = b_\pm$.  Here there is a single component to the conserved current, which we denote $J_\triangle$, and the Ward identity for charge conservation becomes \begin{equation}
    \partial_t\rho + \frac{1}{4}\left(3\partial_x^2 - \partial_y^2\right)  \partial_y J_\triangle = 0. \label{trianglesubsystem}
\end{equation}The derivation is provided in Appendix \ref{app:subsystem}. Consistency with the fluctuation-dissipation theorem demands that $J_\triangle=-C\left(3\partial_x^2 - \partial_y^2\right) \partial_y \rho$, which leads to the very peculiar subdiffusive decay rate for charge at wave number $k$: $\Gamma \propto k_y^2(3k_x^2-k_y^2)^2$.  \revision{As before, these subdiffusive theories are robust, and all nonlinear corrections to the equations of motion are formally irrelevant.}

One reason why the Ward identities are sensitive to the choice of lattice is that in conventional hydrodynamics, to lowest order in gradients, diffusion on square or triangular lattices is rotation invariant, simply because there is not any second-rank tensor that is invariant under the point group symmetry of the lattice.
In fracton hydrodynamics, the extra conservation laws kill the lowest order terms in the derivative expansion, which now starts with higher derivative terms, proportional to non-trivial higher rank tensors that are invariant under the point group.  With subsystem symmetries, the lattice also changes which global charges are conserved, leading to fourth-order subdiffusion for the square lattice versus sixth-order for the triangular lattice.  

The higher dimensional analogues of fracton fluids with subsystem symmetry are straightforward.  In three dimensions, we may consider a theory where charge is conserved on every line of a cubic lattice.  The resulting theory (in the absence of sources) is \begin{equation}
    \partial_t \rho = C \partial_x^2\partial_y^2\partial_z^2\rho. \label{cubesubsystemline}
\end{equation}
If charge is only conserved on every plane, assuming cubic symmetry, the equation becomes \begin{equation}
    \partial_t\rho = -C^\prime \left(\partial_x^2\partial_y^2 + \partial_x^2\partial_z^2 + \partial_y^2\partial_z^2\right)\rho, \label{cubesubsystemplane}
\end{equation}
which has two fewer derivatives. We note that dynamics on square and cubic lattices with charge conservation along lines or planes was studied in \cite{IaconisVijayNandkishore}, where results consistent with (\ref{squaresubsystem}), (\ref{cubesubsystemline}) and (\ref{cubesubsystemplane}) were seen. We have explained these results from a universal hydrodynamic perspective that is independent of microscopic details, and have also provided a new prediction (\ref{trianglesubsystem}) for subdiffusion on triangular lattices. 

\section{Experimental implications}
Our results have direct implications for experiments on constrained quantum dynamics.  As an example, a recent experiment \cite{Guardado_Sanchez_2020} studied thermalization in a cold atomic gas in a tilted optical lattice, where they found that for sufficiently strong tilt, atomic number density modulations of wavelength $\lambda$ relaxed on time scale $\tau \propto \lambda^4$.   At first glance, this experiment seems unconnected to our discussion, since the tilt is not strong enough to enforce dipole conservation on the lattice scale, since the experiment has energy conservation (which our discussion has henceforth neglected), and finally because the experiment has no microscopic fracton excitations. Indeed, within \cite{Guardado_Sanchez_2020} the data was explained in terms of a subtle interplay of two diffusive modes (number and energy).   Nevertheless, the relaxation of the long wavelength number density modulations is described by the same hydrodynamic universality class as a fluid of fractons with local dipole conservation.  Because the experiment conserves energy, in the presence of a non-zero tilt there is an emergent dipole conservation law on hydrodynamic length scales: assuming microscopic energy scale $U$ and external force $F$, a clump of atoms cannot simply diffuse a distance $L$ if $U\ll FL$.  
Once the tilt is applied to the lattice, the diffusive modes of atom number and energy morph on long length scales into \emph{one} subdiffusive hydrodynamic mode and one ``quasihydrodynamic" \cite{Grozdanov:2018fic} mode which decays at a finite rate.  An explicit derivation of (\ref{eq:dipolecons}) and (\ref{eq:Jijfinal}) in this model are provided in Appendix \ref{app:experiment}.  The genuine hydrodynamic limit of this theory is identical to that of a fluid of fractons with dipole conservation, since hydrodynamics is an effective theory and depends only on (emergent) symmetries.   Therefore, our framework naturally explains the observed subdiffusive relaxation $\tau\propto\lambda^4$ on scales $\lambda \gg U/F$.

A natural extension of this work is to study the dynamics of an atomic gas trapped in an optical lattice which is in turn placed in a strong harmonic trap.  For very strong trap strengths, the harmonic potential will lead to an emergent fluid with local \emph{quadrupole conservation} \cite{khemanihermelenandkishore}: see Appendix \ref{app:experiment}.   We predict that charge density modulations in this trapped optical lattice will relax even more slowly:  $\tau \propto \lambda^6$.  This result may be naturally tested in near term ultracold atom experiments.  

\section{Magnetic fields}
Another application of our formalism is to charged two-dimensional fluids in a background magnetic field of strength $B$. For simplicity, we assume Galilean-invariance (though the calculation can be generalized).  It is known \cite{Hartnoll:2007ih} that the sound mode and diffusion mode for transverse momentum morph into cyclotron modes dominated by the momentum density (which is no longer conserved due to the magnetic field) and a subdiffusive mode describing charge relaxation, obeying $\tau \propto \lambda^4$.  We can immediately understand this subdiffusion as arising from an emergent dipole conservation.  In the presence of a background magnetic field, the conserved canonical momenta are $P_{\mathrm{fluid},x} + B Y_{\mathrm{d}}$ and $P_{\mathrm{fluid},y} - B X_{\mathrm{d}}$ where $X_{\mathrm{d}},Y_{\mathrm{d}}$ denote the total dipole moment of the fluid and $P_{\mathrm{fluid}}$ denotes the physical momentum density.  As in the tilted optical lattice, on large wavelengths, the dipole moments dominate the conservation of canonical momentum, and there is a single hydrodynamic subdiffusive mode associated with the relaxation of charge.  A complementary discussion about fracton-like dynamics in a similar system is found in \cite{doshi2020vortices}.

\section{Long-range interactions}
Atomic quantum simulators ranging from polar molecules \cite{junye} to Rydberg atoms \cite{RevModPhys.82.2313} or trapped ion crystals \cite{britton} consist of degrees of freedom which exhibit long-range interactions: clusters of particles, where no two particles are separated by a distance greater than $r$, have interaction energies $E(r) \propto r^{-\alpha}$.   Below what $\alpha$ does  hydrodynamics qualitatively break down \cite{Schuckert:2019dqw}?  
In the presence of $n$-pole subdiffusion, a density fluctuation will travel a distance $r\sim t^{1/(2+2n)}$ in time $t$.  Using Fermi's golden rule, we estimate that in the same time $t$, the typical distance that the charge might jump using a long-ranged interaction is given by $t\sim r^{2\alpha-d}$ (the factor of $r^{2\alpha}$ comes from squaring the matrix elements in the transition rate estimate; the factor of $r^{-d}$ comes from integrating over all possible sites to jump to).  The jumps due to long-range interactions spread the charge as fast (or faster) than subdiffusion when $\alpha \le n+1+\frac{d}{2}$.  When $\alpha>n+1+\frac{d}{2}$, long-ranged interactions do not destroy subdiffusion.

Our argument further implies that any exponentially suppressed long-ranged virtual processes permitted in the tilted lattice experiment of \cite{Guardado_Sanchez_2020} do not break the subdiffusive dynamics, even in the thermodynamic limit.  A more interesting proposal is to repeat the experiment using degrees of freedom with microscopic dipole-dipole interactions ($\alpha=3$).  While the untilted lattice exhibits diffusive charge dynamics in three (and below) dimensions, a tilted system with approximate dipole conservation will \emph{not} be as subdiffusive as a system with local interactions in three dimensions.

\section{Outlook}
The past decade has seen a resurgence of study into the hydrodynamics of quantum fluids, which are usually described by the Navier-Stokes equations or mild modifications thereof.  Our study of fracton fluids has revealed infinitely many hitherto undiscovered universality classes of hydrodynamic behavior with clear experimental signatures in both static and dynamical transport that are qualitatively distinct from conventional Navier-Stokes hydrodynamics.  Since hydrodynamic equations are ultimately classical, it may also be possible to mimic these effects using engineered active matter \cite{Nash14495}.  We look forward to the future theoretical and experimental efforts to uncover, classify and realize microscopically the many universality classes of fracton hydrodynamics.

\section*{Acknowledgements}
We thank Paolo Glorioso, David Huse and Alan Morningstar for useful discussions. RMN would like to acknowledge prior collaborations on related topics with B. Andrei Bernevig, Michael Hermele, Sanjay Moudgalya, Shriya Pai, Abhinav Prem, Michael Pretko, Nicolas Regnault and especially Vedika Khemani. RMN was supported in part by the Air Force Office of Scientific Research under award number FA9550-17-1-0183. RMN also acknowledges the support of the Alfred P. Sloan Foundation through a Sloan Research Fellowship. AG was supported by Brown University.  AL was supported by the University of Colorado.

\emph{Note Added.}---After our work appeared on the arXiv, other groups reported similar predictions for subdiffusion in fluids with conserved multipole moments \cite{morningstar2020kineticallyconstrained,feldmeier2020anomalous,zhang2020universal}.  L. Radzihovsky (unpublished) has derived subdiffusion with conserved dipole moment using fracton-elasticity duality.

\onecolumngrid

\begin{appendix}
\section{Effective field theory of hydrodynamics with multipole conservation}
\label{app:multipole}
\subsection{Review of diffusion}
First we review the effective theory of hydrodynamics developed in \cite{Crossley:2015evo}, in the linear response regime, and for the simple case of a single conserved U(1) charge.  After thoroughly summarizing the earlier procedure, it will be immediate to extend the work to phases of matter with fracton excitations.  Hence, we begin by studying a theory that only conserves the ``0-pole" charge.  Consider the generating functional of hydrodynamic correlation functions \begin{equation}
    Z[A_t, A_i] = \left\langle \exp\left[\mathrm{i}\int \mathrm{d}^{d+1}x ( A_{t}(x_1) \rho(x_1) + A_{i}(x_1)J^{i}(x_1)) -( A_{t}(x_2) \rho(x_2) + A_{i}(x_2)J^{i}(x_2)) \right]\right\rangle, \label{eq:ZappA}
\end{equation}
where $J^i$ here denotes the ordinary charge current, $x_1$ denotes a spacetime point on the forward time contour, and $x_2$ denotes a point on the backwards time contour.  We have defined this action on a Schwinger-Keldysh contour because our ultimate goal is to derive a dissipative effective theory: hydrodynamics. A contour that runs both forward and backward in time is required to obtain the correct operator orderings to study hydrodynamic correlation functions.

Due to long wavelength hydrodynamic fluctuations which have been integrated out, $Z[A_t,A_i]$ is highly non-local. The authors of \cite{Crossley:2015evo} argue that one must ``integrate in" the hydrodynamic fluctuations. The hydrodynamic degree of freedom in the effective action corresponds to a local U(1) phase rotation $\phi$ in each fluid element: \begin{equation}
    A_t \rightarrow A_t + \partial_t \phi, \;\;\; A_i \rightarrow A_i + \partial_i \phi.
\end{equation}
So we postulate that \begin{equation}
    Z[A_t,A_i] = \int \mathrm{D}\phi \; \mathrm{e}^{\mathrm{i} I[B_t, B_i]}
\end{equation}
where \begin{equation}
    B_t = A_t + \partial_t \phi, \;\;\;\; B_i = A_i + \partial_i \phi .
\end{equation}
It is useful to think of two fields living on a single time contour, instead of one field living on a two-sided contour.  So we define\begin{equation}
    B_{\mu +} (x) = \frac{B_\mu(x_1) + B_\mu(x_2)}{2}, \;\;\;\; B_{\mu -} (x) = B_\mu(x_1) - B_\mu(x_2).
\end{equation}
The $B_-$ field corresponds to the stochastic noise field, while the $B_+$ field corresponds to the hydrodynamic mode, in a way that we will clarify shortly.

We now wish to build up $I$ using the principles of effective field theory.  For the purposes of this paper, we restrict ourselves to quadratic actions.  There are a number of symmetries that we must impose, which we list here (see \cite{Crossley:2015evo} for the justification of these facts): \begin{enumerate}
    \item \emph{Spacetime symmetries}: $I$ must be independent of spacetime position $x$ and all spatial indices must be contracted.  We assume the action is local.
     \item \emph{Reflection symmetry:} $I[B_+, B_-] = -I[B_+, -B_-]^*$: switching the order of the contours means that $\mathrm{i}I$ is complex conjugated.
    \item \emph{Unitarity:} $I[B_- = 0] = 0$.  All terms in $I$ must have at least one $-$ field.
    \item \emph{Fluid phase relabeling}: The initial choice of fluid phase at each point may be freely chosen at the initial time $t=0$, so $I$ must be invariant under $\phi_+ \rightarrow \phi_+ + \lambda(x_i)$.
   \revision{     \item \emph{Kubo-Martin-Schwinger (KMS) symmetry}: In a system at finite temperature $T$, suppose that our quadratic action is of the form \begin{equation}
            I = \int \mathrm{d}^{d+1}x\mathrm{d}^{d+1}x^\prime \left( \frac{\mathrm{i}}{2} G_{\alpha\beta}(x,x^\prime) B_{\alpha-}(x) B_{\beta-}(x^\prime) + K_{\alpha\beta}(x,x^\prime)B_{\alpha-}(x) B_{\beta+}(x^\prime) \right)
        \end{equation}
        where $\alpha,\beta$ correspond to different field indices (for example, we might take $B_\alpha$ to be either $B_t$ and $B_i$).  Above we let $x$ include both the $d$ spatial coordinates and the time coordinate $t$.  Schematically,  \begin{equation}
            G_{\alpha\beta}(x,x^\prime) \sim \langle \lbrace O_\alpha(x), O_\beta(x^\prime)\rbrace \rangle, \;\;\;  K_{\alpha\beta}(x,x^\prime) \sim  \mathrm{\Theta}(t^\prime-t) \mathrm{i} \langle [O_\alpha(x),O_\beta(x^\prime)]\rangle. \label{eq:GKdef}
        \end{equation}where $O_\alpha$ represents the hydrodynamic operator which couples to $B_\alpha$ in (\ref{eq:ZappA}) -- see \cite{Crossley:2015evo} for details.  Note that (\ref{eq:GKdef}), together with unitarity, time reversal symmetry and emergent space translation symmetry, also implies that \begin{equation}
            K_{\alpha\beta}(x,x^\prime) = K_{\alpha\beta}(x-x^\prime) = K_{\beta\alpha}(x^\prime-x). \label{eq:Kindex}
        \end{equation}   By translation invariance, we may Fourier transform $G$ and $K$.  The Fourier transforms obey \begin{equation}
            G_{\alpha\beta}(k) = G_{\beta\alpha}(-k) = -\mathrm{i}\frac{T}{\omega} \left(K_{\alpha\beta}(k) - K_{\alpha\beta}(k)^*\right).  \label{eq:KMS}
        \end{equation}
        for $\omega \ll T$ (the hydrodynamic limit).   Note that only the last step, (\ref{eq:KMS}), actually relies on energy conservation.  Hence, many of the results we derive below are also relevant for theories without energy conservation, but with an emergent or averaged time translation invariance and time reversal symmetry.
        }
\end{enumerate}

\revision{For convenience, we will assume KMS symmetry for the moment, as this is the context in which the effective theories above are best understood.  However, we will also argue at the end that our qualitative conclusions are not sensitive to this result.} We may now write down the dominant terms in $I$ compatible with the 5 symmetries above.  If $a$, $b$ and $c$ are real constants, then 
\begin{equation}
I = \int \mathrm{d}^{d+1}x \left[ \frac{\mathrm{i}}{2}\left(a B_{t-}^2 + 2bT B_{i-}^2 \right) + c B_{t-}B_{t+} - bB_{i-}\partial_t B_{i+} + \cdots  \right] \label{eq:fickaction}
\end{equation}
where $\cdots$ contains higher derivative contributions. $c>0$ is required by thermodynamic consistency, and this choice of sign is further consistent with $\phi$ having a positive-signed ``kinetic" term in a Lagrangian.  $a>0$ and $b>0$ are both required by the the fact that the path integral weight $\mathrm{e}^{\mathrm{i}I}$ cannot diverge.   We also emphasize that we cannot include a term of the form $(\partial_i B_{i-}) B_{t+}$ in $I$, since (\ref{eq:KMS}) would lead to another term $(\partial_i B_{i+}) B_{t-} $ which is forbidden by the phase relabeling symmetry.

To find the classical hydrodynamic equations in the absence of noise, we now evaluate \begin{equation}
    \left.\frac{\delta I}{\delta \phi_-}\right|_{B_-=0} = 0 =  -c\partial_t B_{t+} + b\partial_i \partial_t B_{i+}. \label{eq:fickEOM}
\end{equation}
At long last, we identify the chemical potential $\mu$, charge density $\rho$, external electric field $E_i$, and diffusion constant $D$ as \begin{equation}
    \mu = B_{t+}, \;\;\;\; \rho = c\mu, \;\;\;\; E_i = \partial_i A_t - \partial_t A_i, \;\;\;\; D = \frac{b}{c}. \label{eq:muidentity}
\end{equation}
The last equality above is the standard Einstein relation.  Combining (\ref{eq:fickEOM}) and (\ref{eq:muidentity}) we find Fick's law of diffusion, \begin{equation}
    \partial_t \rho = b \partial_i \left(\partial_i \mu - E_i \right) = D \nabla^2 \rho - b \partial_i E_i.
\end{equation}

\subsection{Effective theories for subdiffusion}

Having carefully derived the hydrodynamic equations with 0-pole conservation, we now turn to the case of $n$-pole conservation.  We assume the same symmetries as before, including rotational invariance for simplicity (this is straightforward to relax).  

We emphasized in the main text that we must no longer think about our hydrodynamic theory as coupling to external sources $A_t$ and $A_i$ (the ordinary U(1) gauge field) -- rather, we must couple to $(A_t,A_{i_1\cdots i_{n+1}})$.  Hence, we must build the action out of \begin{equation}
    B_t = A_t + \partial_t \phi, \;\;\;\; B_{i_1\cdots i_{n+1}} = A_{i_1\cdots i_{n+1}} + \partial_{i_1}\cdots \partial_{i_{n+1}} \phi.
\end{equation}
All of the possible spatial index structures in $A_{i_1\cdots i_{n+1}}$ have  enough derivatives to kill \emph{all possible} polynomial shifts in $\phi$ of order $n$.  This encodes the multipole algebra \cite{gromov2018towards} in our hydrodynamic effective action.  After all, another way to interpret the mixed rank gauge field is that it is the minimal object which is guaranteed to vanish under all of the shift symmetries of the theory (corresponding to adjusting any or all of the conserved quantities, which in this case are the multipole moments).

The simplest action that we can write down is analogous to (\ref{eq:fickaction}):
\begin{equation}
I = \int \mathrm{d}^{d+1}x \left[ \frac{\mathrm{i}}{2}a B_{t-}^2 + c B_{t-}B_{t+} - b_{i_1\cdots i_{n+1}j_1\cdots j_{n+1}}B_{i_1\cdots i_{n+1}-}(\partial_t B_{j_1\cdots j_{n+1}+} - \mathrm{i}TB_{j_1\cdots j_{n+1}-}) + \cdots  \right], \label{eq:subdiffaction}
\end{equation}
where again thermodynamic consistency and bounded noise spectrum imply that $a,b,c>0$. Here \begin{equation}
    b_{i_1\cdots i_{n+1}j_1\cdots j_{n+1}}=b_{j_1\cdots j_{n+1}i_1\cdots i_{n+1}} = b_{(i_1\cdots i_{n+1})j_1\cdots j_{n+1}} \label{eq:btensor}
\end{equation} is a tensor structure built up out of Kronecker delta symbols: e.g. when $n=1$, $b_{ijkl} = b_1\delta_{ij}\delta_{kl}+ b_2 (\delta_{ik}\delta_{jl}+\delta_{il}\delta_{jk})$.    In general, (\ref{eq:btensor}) allows $\lceil 1+\frac{n+1}{2}\rceil$ distinct possible terms in the $b$ tensor.  There are two points worth emphasizing here.  For simplicity take $n=1$ (dipole conservation), though both issues generalize.  (\emph{1}) Consider temporarily the case of dipole conservation ($n=1$).   \revision{The reason we cannot write down $B_{ii-}B_{t+}$ is that consistency with (\ref{eq:GKdef}) and (\ref{eq:Kindex}) would demand a term $B_{ii+}B_{t-}$ which is not consistent with phase relabeling symmetry.}  We must have a $\partial_t$ in every term with $B_{ij+}$.  (\emph{2}) While we can indeed write down terms of the form $B_{t+}\partial_t B_{ii-} + B_{t-}\partial_t B_{ii+}$, and these do strictly speaking have fewer derivatives, since the terms in the equations of motion with the fewest \emph{time derivatives} will be of the form $(\partial_t\mu  + \partial_x^4 \mu ) = 0 $ (schematically), every time derivative counts for 4 spatial derivatives; hence, $B_{t-}\partial_t B_{ii+}$ will lead to a subleading correction to hydrodynamics when compared to $B_{ij-}\partial_tB_{ij+}$.

We compute the equations of motion analogously to before.  Defining \begin{equation}
     \mu = B_{t+}, \;\;\;\; \rho = c\mu, \;\;\;\; E_{i_1\cdots i_{n+1}} = (-1)^{1+n} \partial_{i_1}\cdots \partial_{i_{n+1}} A_t - \partial_t A_{i_1\cdots i_{n+1}},
\end{equation}
and varying $I$ with respect to $\phi_-$, we obtain \begin{equation}
    \partial_t \rho =  b_{i_1\cdots i_{n+1}j_1\cdots j_{n+1}}\partial_{i_1}\cdots \partial_{i_{n+1}}\left(\partial_{j_1}\cdots \partial_{j_{n+1}}\mu - E_{j_1\cdots j_{n+1}}\right).
\end{equation}

\subsection{Scaling dimensions}
\revision{
Using the effective action (\ref{eq:subdiffaction}), it is straightforward to determine the various operator dimensions of the effective theory and to confirm that all deformations are irrelevant.   As we have already seen, the linearized stochastic subdiffusive equations are controlled by the $c$ and $b$ terms in (\ref{eq:subdiffaction}).  Letting $[O]$ denote the scaling dimension of operator $O$, and fixing $[x]=-1$ by definition, we find the following set of equations from the $c$ term and the $b$ terms respectively: \begin{subequations}\begin{align}
   d - [t] &= -2[t] + [\phi_+] + [\phi_-], \\
   d - [t] &= -[t] + 2(n+1) + [\phi_+] + [\phi_-], \\
   d - [t] &= 2(n+1) + 2[\phi_-],
\end{align}\end{subequations}
which can be straightforwardly solved to give \begin{subequations}\begin{align}
    [t] &= -2(n+1), \\
    [\phi_-] &= \frac{d}{2}, \\
    [\phi_+] &= \frac{d}{2}-2(n+1).
\end{align}\end{subequations}
In other words, in our subdiffusive fixed point, the parameters $b$ and $c$ must be ``dimensionless" coupling constants under renormalization group flow to longer length scales.  We emphasize again that in hydrodynamics, this flow does \emph{not} correspond to reducing the temperature, which plays the role of an ``ultraviolet" scale.  

A priori, it appears that $\phi_+$ can be relevant.  However, the shift symmetry in the effective action mandates that $I$ can only depend on $\mu = \partial_t \phi_+$, which has dimension \begin{equation}
    [\mu] = [\partial_t \phi_+] = [\phi_+]-[t] = \frac{d}{2}.
\end{equation}
Hence in all physical spatial dimensions $d=1,2,3,\ldots$, all further terms that show up in (\ref{eq:subdiffaction}) beyond $b$ and $c$ are irrelevant corrections (including the $a$ term).  This demonstrates that our subdiffusive theories are universal and completely stable fixed points.

}

\subsection{Beyond KMS}

\revision{Finally, let us return to the matter of KMS symmetry.  If we wish to study emergent hydrodynamics in any system without energy conservation, it is not appropriate to enforce such a symmetry.  Let us ask what changes need to be made.  In the enumerated list of the first subsection, only point 5 needs to be relaxed.  Assuming that $I$ must still be built out of St\"uckleberg fields, then the most general action we can write down is \begin{equation}
    I = \int \mathrm{d}^{d+1}x \left[ \frac{\mathrm{i}}{2}a B_{t-}^2 + c B_{t-}B_{t+} - b_{i_1\cdots i_{n+1}j_1\cdots j_{n+1}}B_{i_1\cdots i_{n+1}-}\partial_t B_{j_1\cdots j_{n+1}+} + \mathrm{i} f B_{i_1\cdots i_{n+1}-}^2 + \cdots  \right]
\end{equation}
where the only change relative to (\ref{eq:subdiffaction}) is that the coefficient $f$ is no longer constrained by a thermal fluctuation dissipation theorem.  The condition that allows us to forbid terms such as $B_{ii-} B_{t+}$ in the dipole conserving theory is (\ref{eq:GKdef}), which is not lost in the absence of energy conservation.  In addition, the effective operator dimensions and equations of motion are unchanged, and thus our conclusions remain relevant for systems without energy conservation.
}

\section{Effective field theory of hydrodynamics with subsystem symmetry}
\label{app:subsystem}

\subsection{2d square lattice}
We will use this as an illustrative example, while the other scenarios discussed in the main text will be fairly straightforward extensions.  The starting point is to begin by thinking about the phase field $\phi$.  The infinite family of conservation laws (\ref{eq:squareconservation}) is summarized by demanding that the action be invariant under \begin{equation}
    \phi_- \rightarrow \phi_- + \lambda(x) + \zeta(y)  \label{eq:squarelatticeshift}
\end{equation}
for arbitrary functions $\lambda$ and $\zeta$. What gauge field $A$ would be compatible with these symmetries?  Clearly we cannot choose $A_{xxxxx}$, for example, since $A_{xxxxx} \rightarrow A_{xxxxx} + \partial_x^5 \phi$ is not invariant under $\phi = x^5$, which is a global symmetry of the problem.   Every ``spatial component" of $A$ must have at least one $x$ and one $y$ index, so that the two separate shifts in (\ref{eq:squarelatticeshift}) are cancelled.   There is a unique simplest tensor structure, $A_{xy}$, which contains the fewest number of derivatives and is invariant.  This explains the gauge field $(A_t,A_{xy})$ described in the main text.

Using the above framework, the effective action $I$ must be built out of the gauge fields \begin{equation}
B_t = A_t + \partial_t \phi , \;\;\;\; B_{xy} = A_{xy} + \partial_x\partial_y\phi.    
\end{equation}
At leading order, the action is \begin{equation}
    I = \int \mathrm{d}^{d+1}x \left[ \frac{\mathrm{i}}{2}\left(a B_{t-}^2 + 2bT B_{xy-}^2 \right) + c B_{t-}B_{t+} - bB_{xy-}\partial_t B_{xy+} + \cdots  \right]
\end{equation}
Note that terms of the form $B_{xy-}\partial_t B_{t+}$ are forbidden because they are not invariant under the point group symmetry of the square lattice, which includes $(x,y)\rightarrow (-x,y)$.  It is straightforward to obtain the fourth order subdiffusion equation of the main text from this effective action.

\subsection{2d triangular lattice}  Let us orient the $(x,y)$ coordinate system such that the edges of the triangular lattice are oriented in the following directions:  $\hat{\mathbf{y}}$, $\pm \frac{\sqrt{3}}{2}\hat{\mathbf{x}} + \frac{1}{2}\hat{\mathbf{y}}$. Now the symmetries of our effective action must include \begin{equation}
    \phi_- \rightarrow \phi_- + \lambda(x) + \zeta\left(\frac{\sqrt{3}y+x}{2}\right)+ \eta\left(\frac{\sqrt{3}y-x}{2}\right)  \label{eq:triangularlatticeshift}
\end{equation}
for arbitrary functions $\lambda$, $\zeta$ and $\eta$.  A $\lambda$-shift corresponds to changing the charge density on every line oriented in the $y$-direction, which is allowed since each is separately conserved; the $\zeta$ and $\eta$ shifts correspond to adjustments in the total charge on the other two ``directions" of the lattice, which also are individually conserved.  

From the form of (\ref{eq:triangularlatticeshift}) it is clear that $\phi$ can only appear in the action in the form $\partial_t \phi$ or $\partial_y (\sqrt{3}\partial_x -y)(\sqrt{3}\partial_x +y)\phi$:  respectively these there spatial derivatives annihilate the $\lambda$, $\zeta$ and $\eta$ terms, while the time derivative annihilates all.  Just as on the square lattice, at least one component orthogonal to each of the lattice directions is mandatory for all spatial components of $A$; we conclude that the effective action will be built out of \begin{equation}
    B_t = A_t + \partial_t \phi , \;\;\;\; B_\triangle = A_\triangle + \partial_y (\sqrt{3}\partial_x -\partial_y)(\sqrt{3}\partial_x +\partial_y) \phi,
\end{equation}
where we have resorted to the label $\triangle$ for the spatial indices of $B$ or $A$ (note that these are single component rank-3 spatial tensors, which transform under the one-dimensional ``spin 3" representation of the dihedral group $\mathrm{D}_{12}$, which represents the point group of the lattice).   The effective action is \begin{equation}
    I = \int \mathrm{d}^{d+1}x \left[ \frac{\mathrm{i}}{2}\left(a B_{t-}^2 + 2bT B_{\triangle-}^2 \right) + c B_{t-}B_{t+} - bB_{\triangle-}\partial_t B_{\triangle+} + \cdots  \right].
\end{equation}
As on the square lattice, the point group symmetry includes parity which demands that two $\triangle$ indices must always come together.  Varying $I$ with respect to $\phi_-$ and neglecting noise leads to (\ref{trianglesubsystem}).

\subsection{3d cubic lattice: charge conserved on lines}  This is a straightforward extension of the 2d square lattice model.  We demand \begin{equation}
    \phi_- \rightarrow \phi_- + \lambda(x,y)+\zeta(y,z)+\eta(z,x)
\end{equation}
for arbitrary functions $\lambda$, $\zeta$ and $\eta$.  The gauge field has one unique spatial component $A_{xyz}$, and we build the action out of \begin{equation}
    B_t = A_t + \partial_t \phi , \;\;\;\; B_{xyz} = A_{xyz} + \partial_x\partial_y\partial_z \phi.
\end{equation}
The effective action which leads to (\ref{cubesubsystemline}) is \begin{equation}
     I = \int \mathrm{d}^{d+1}x \left[ \frac{\mathrm{i}}{2}\left(a B_{t-}^2 + 2bT B_{xyz-}^2 \right) + c B_{t-}B_{t+} - bB_{xyz-}\partial_t B_{xyz+} + \cdots  \right].
\end{equation}

\subsection{3d cubic lattice: charge conserved on planes} Now we demand that \begin{equation}
    \phi_- \rightarrow \phi_- + \lambda(x)+\zeta(y)+\eta(z)
\end{equation}
for arbitrary functions $\lambda$, $\zeta$ and $\eta$.  The gauge field has three different two-index components: $A_{xy},A_{yz},A_{zx}$, since any function of two distinct coordinates is not included in the shift symmetry above.  Hence, we build the action out of \begin{equation}
    B_t = A_t + \partial_t \phi , \;\;\;\; B_{xy} = A_{xy} + \partial_x\partial_y \phi, \;\;\;\; B_{yz} = A_{yz} + \partial_y\partial_z \phi, \;\;\;\; B_{zx} = A_{zx} + \partial_z\partial_x \phi.
\end{equation}
The effective action which is invariant under the point group of the lattice, which includes inversion as well as rotating among the $x$, $y$ and $z$ directions, leads to (\ref{cubesubsystemplane}): \begin{align}
     I &= \int \mathrm{d}^{d+1}x \left[ \frac{\mathrm{i}}{2}\left(a B_{t-}^2 + 2bT \left(B_{xy-}^2 +B_{yz-}^2 +B_{zx-}^2 \right)  \right) + c B_{t-}B_{t+} \right. \notag \\
     &\left. - b\left(B_{xy-}\partial_t B_{xy+} + B_{yz-}\partial_t B_{yz+}+B_{zx-}\partial_t B_{zx+} \right) + \cdots  \right].
\end{align}

\section{Emergent multipole-conserving hydrodynamics in a polynomial potential}
\label{app:experiment}
In this appendix, we consider one dimensional fluids, and describe how subdiffusive relaxation generically arises in fluids with charge and energy conservation, but contained inside higher order polynomial potentials: \begin{equation}
    V_{\mathrm{pot}}(x) = \frac{K}{n}x^n.
\end{equation}

\subsection{Linear potentials}  We begin with a pedagogically simplified version of the model presented in \cite{Guardado_Sanchez_2020} to show the emergence of our dipole-conserving hydrodynamic subdiffusion in their model.  This corresponds to choosing $n=1$ and $K=-F$.  Let $\rho$ denote the atomic number density, and $e - Fx\rho$ denote the total energy density: note that $e$ will not count the ``tilt energy" $Fx\rho$.  The equations for $\rho$ and $e$ may be written as \cite{Guardado_Sanchez_2020} \begin{equation}
    \partial_t \rho + \partial_x J_\rho = 0 , \;\;\;\; \partial_t e + \partial_x J_e = FJ_\rho. \label{eq:rhoe}
\end{equation}
Here $J_\rho$ and $J_e$ denote number and energy flux respectively.  We assume that the thermal energy flux and the atomic number flux may be approximated as \begin{equation}
    J_e \approx -M_e \partial_x e + \cdots, \;\;\;\; J_\rho \approx -M_\rho (s_\rho \partial_x \rho + s_e F e) + \cdots. \label{eq:Je}
\end{equation}
Here $M_e$ and $M_\rho$ are dissipative coefficients, while $s_\rho$ and $s_e$ are thermodynamic coefficients.  We have neglected some cross-terms in (\ref{eq:Je}), but their inclusion does not change the qualitative physics;  see \cite{Guardado_Sanchez_2020} for the general case.

\revision{Combining (\ref{eq:rhoe}) with (\ref{eq:Je}) we find that \begin{equation}
    \partial_t e + F^2 M_\rho s_ee = \cdots,
\end{equation}
which shows that the thermal energy density $e$ is not conserved in the presence of a tilt.}  Hence, the slowest degrees of freedom in the system will exhibit ultra-slow number diffusion with \begin{equation}
    FJ_\rho = \partial_x J_e + \cdots . \label{eq:FJrho}
\end{equation}
The $\cdots$ above will contain higher derivative corrections and may be ignored in the hydrodynamic limit.  Upon defining the dipole current \begin{equation}
    J_{\mathrm{dipole}} = \frac{J_e}{F},
\end{equation}
which can be understood as a consequence of macroscopic dipole flux necessarily arising through the relaxation of the tilt energy, we obtain the hydrodynamic equation of motion (\ref{eq:dipolecons}) for a fluid with an \emph{emergent local dipole conservation}: \begin{equation}
    \partial_t \rho + \partial_x^2 J_{\mathrm{dipole}} = 0. \label{eq:experimentdipole}
\end{equation}
To check that $J_{\mathrm{dipole}} \propto \partial_x^2 \rho$, we recall (\ref{eq:Je}).  If $J_\rho$ is a total derivative, then $s_e Fe = -s_\rho \partial_x \rho + \cdots$, where $\cdots$ again represents higher derivative corrections.  Therefore, \begin{equation}
    J_{\mathrm{dipole}} \approx \frac{M_e s_\rho}{F^2}\partial_x^2 \rho, \label{eq:experimentcurrent}
\end{equation}
consistent with both our generic framework and the result of \cite{Guardado_Sanchez_2020}.

Combining (\ref{eq:experimentdipole}) and (\ref{eq:experimentcurrent}), we obtain that one  quasinormal mode in this model hence obeys has the following relation: \begin{equation}
    \omega = -\mathrm{i}\frac{M_e s_\rho}{F^2} k^4+\cdots.
\end{equation}The second quasinormal mode in this theory is not a genuine hydrodynamic mode, in that its lifetime remains finite as the wavelength $\lambda$ of fluctuations diverges.  In this other mode, $J_\rho$ does not vanish at leading order, and the energy conservation equation can be approximated by \begin{equation}
    \partial_t e = -M_\rho F(s_\rho \partial_x \rho + s_e Fe).
\end{equation}  
We conclude that the second quasinormal mode obeys
\begin{equation}
\omega = -\mathrm{i}F^2 M_\rho s_e+\cdots  .
\end{equation}
In the language of \cite{Grozdanov:2018fic}, this second mode is called ``quasihydrodynamic" -- it has a parametrically slow decay rate in the limit $F\rightarrow 0$; however, because $\omega(k=0)\ne 0$, it is not a genuine hydrodynamic mode.   The true hydrodynamic limit of the quasihydrodynamic model of \cite{Guardado_Sanchez_2020} is, therefore, the same hydrodynamics as models with locally conserved dipole moment.

\subsection{Higher order potentials} We now generalize the above discussion  to $n>1$.  (\ref{eq:rhoe}) and (\ref{eq:Je}) immediately generalize upon making the force $x$-dependent: \begin{equation}
    F(x) = -kx^{n-1}.
\end{equation}
The remainder of the argument follows through.  Since the thermal energy $e$ is not conserved, the slowest dynamics will only involve $\rho$.  Combining (\ref{eq:rhoe}), (\ref{eq:Je}) and (\ref{eq:FJrho}), we find \begin{equation}
    \partial_t \rho \approx -\partial_x \frac{M_e}{F(x)} \partial_x^2 \frac{s_\rho}{F(x)}\partial_x \rho. \label{eq:higherorderrho}
\end{equation}
Now the force is no longer homogeneous, and so plane wave solutions do not exist.  Nevertheless, we may still look for normal modes of the form \begin{equation}
    \rho(x,t) = \rho_0\left(\frac{x}{\lambda}\right)\mathrm{e}^{-\Gamma t},
\end{equation}
for relaxation rate $\Gamma$, unknown function $\rho_0$, and arbitrary wavelength $\lambda$.  Indeed, we observe that \begin{equation}
    \Gamma =  \frac{1}{\lambda^{2n+2}}\frac{M_e s_\rho}{K^2},
\end{equation}
and that upon making these substitutions into (\ref{eq:higherorderrho}), we find \begin{equation}
  \rho_0(z) = \partial_z \frac{1}{z^{n-1}}\partial_z^2 \frac{1}{z^{n-1}}\partial_z \rho_0(z),
\end{equation}
which means that normal modes exist for all wavelengths $\lambda$.  We find that $\rho_0$ is given by \begin{equation}
    \rho(z) = C_1 z^{n/2} \mathrm{J}_{-n/(n+1)}\left(\frac{2z^{(n+1)/2}}{n+1}\right) + C_2z^{n/2}\mathrm{J}_{n/(n+1)}\left(\frac{2z^{(n+1)/2}}{n+1}\right)
\end{equation}
where $C_1$ and $C_2$ are arbitrary constants and $\mathrm{J}_\alpha$ denotes the Bessel function of order $\alpha$.   These solutions are well-behaved as a function of $x$ and replace the ordinary plane wave solutions that we find in the case $n=1$.  Although we no longer find ordinary plane wave solutions, the relationship between relaxation rate $\Gamma$ and wavelength $\lambda$ does not change.
\end{appendix}

\bibliography{bibliography5}

\end{document}